\documentclass[conference]{IEEEtran}
\usepackage{lipsum}
\makeatletter
\def\footnoterule{\relax%
  \kern-5pt
  \hbox to \columnwidth{\hfill\vrule width \columnwidth height 0.4pt\hfill}
  \kern4.6pt}
\makeatother                 

\usepackage{cite}         
\usepackage{amssymb}
\usepackage{lettrine}                              

\usepackage{makecell}
\usepackage{graphicx}
\usepackage{float}
\usepackage{amsmath}
\usepackage{tabularx}
\usepackage{multirow}
\makeatletter
\let\NAT@parse\undefined
\makeatother
\usepackage{hyperref}

\IEEEoverridecommandlockouts                              

\title{\huge On the Use of Audio Fingerprinting Features for Speech Enhancement with Generative Adversarial Network}

\author{Farnood Faraji$^{a,*}$, Yazid Attabi$^a$, Benoit Champagne$^a$ and Wei-Ping Zhu$^b$  \\
$^a$Department of Electrical and Computer Engineering, McGill University, Montreal, QC, Canada\\
$^b$Department of Electrical and Computer Engineering, Concordia University, Montreal, QC, Canada
\thanks{Funding for this work was provided by a grant from NSERC Canada with industrial sponsor Microsemi Corporation.}
\thanks{$^*$Corresponding Author: farnood.faraji@mail.mcgill.ca}
} 

\begin{document}
\maketitle
\thispagestyle{empty}
\pagestyle{empty}
\setcounter{page}{1}
\pagenumbering{arabic}

\begin{abstract}
The advent of learning-based methods in speech enhancement has revived the need for robust and reliable training features that can compactly represent speech signals while preserving their vital information. 
Time-frequency domain features, such as the Short-Term Fourier Transform (STFT) and Mel-Frequency Cepstral Coefficients (MFCC), are preferred in many approaches.
While the MFCC provide for a compact representation, they ignore the dynamics and distribution of energy in each mel-scale subband.
In this work, a speech enhancement system based on Generative Adversarial Network (GAN) is implemented and tested with a combination of Audio FingerPrinting (AFP) features obtained from the MFCC and the Normalized Spectral Subband Centroids (NSSC). 
The NSSC capture the locations of speech formants and complement the MFCC in a crucial way.
In experiments with diverse speakers and noise types, GAN-based speech enhancement with the proposed AFP feature combination achieves the best objective performance 
while reducing 
memory requirements and training time.
\end{abstract}
\section{Introduction}
\par
Speech enhancement aims to isolate a desired speech signal from the additive background noise, and increase the quality or intelligibility of the processed speech \cite{benesty2005}.
In the past decade, due to important theoretical advances, faster and cheaper computational resources, and the availability of large recorded data set for training, neural networks have been applied successfully to a variety of non-linear mapping problems, including speech enhancement.
For instance, \cite{xu2015} proposes a supervised speech enhancement system based on Deep Neural Network (DNN) that can outperform the conventional methods.
\par The Generative Adversarial Network (GAN) aims to generate more realistic output patterns that exhibit characteristics closer to the real data \cite{goodfellow2014}.
Adversarial training can also be employed in the field of speech enhancement. 
Proposed by \cite{pascual2017}, Speech Enhancement GAN (SEGAN) works in time-domain and uses a one dimensional Convolutional Neural Network (CNN). 
A similar architecture is investigated by \cite{donahue2017} using Short-Term Fourier Transform (STFT) features. Studies by \cite{soni2018} and \cite{pandey2018} use Gammatone spectrum and STFT features, respectively, and propose modified training targets. 
Neural network systems require substantial training data to give the best performance.
Thus, having a reliable feature set which reduces memory requirements and training time is an important asset, especially for embedded systems and real-time applications. Speech enhancement with GAN can work in both time \cite{pascual2017,abdul2019} and frequency domains \cite{donahue2017,soni2018,pandey2018}. However, these works indicate that frequency-domain features have a clear advantage over the former, especially in terms of measures like Perceptual Evaluation of Speech Quality (PESQ) \cite{abdul2019}.

\par Frequency-domain features such as STFT, Gammatone spectrum and Mel-Frequency Cepstral Coefficients (MFCC) have been used frequently. In addition, a combination of STFT with MFCC is employed in \cite{ribas2019} for training wide residual networks for speech enhancement.
Compared to STFT, filter-based features like MFCC exhibit reduced dimensionality and are more suitable for learning algorithms, 
as they can reduce memory and computational requirements
while maintaining comparable level of performance
\cite{pandey2018,razani2017,chen2014}.
MFCC belong to a larger family of so-called Audio Fingerprinting (AFP) features, which include the Spectral Subband Centroids (SSC) and Spectral Energy Peaks (SEP), and are used to compress data and extract essential patterns in audio frames \cite{duong2015}. 
\par
The MFCC are computed by applying the Discrete Cosine Transform (DCT) to a set of weighted subband energies obtained from a Mel-spaced filterbank. 
The filter-based energy computation of this process ignores important information about the audio signal in each subband, such as the locations of energy peaks corresponding to speech formants.
The SSC introduced by Paliwal \cite{paliwal1998}, provides crucial information about the centroid frequency in each subband, which has proven to be of great value in several applications. The SSC have been successfully employed in speech recognition, speaker identification and music classification, with non-learning or dictionary-based systems \cite{poh2004,nicolson2018,Kinnunen2007}. Besides a combination of MFCC and SSC was proposed for speaker authentication with non-learning methods in \cite{thian2004}.

\par In this paper, a state-of-art speech enhancement system based on GAN is implemented to predict the Ideal Ratio Mask (IRM) of the noisy speech, using a compact set of features obtained from the combination of MFCC, Normalized SSC (NSSC) and their time differences (i.e. delta versions). 
The performance of the resulting systems is evaluated by means of standard objective measures, and compared to that of other possible combinations of features, including the STFT coefficients.
Our results show that the proposed combination of AFP features based on MFCC and NSSC can achieve best (or near best) performance under a wide range of SNR, while significantly reducing memory requirements and training time.


\section{Generative Adversarial Network}
\label{sec:gan}
\par 
GANs are generative models designed to map noisy sample vectors $\mathbf{z}$ from a prior distribution into outputs that resemble those generated from the real (i.e. actual) data distribution. 
To achieve this, a generator (G) learns to effectively imitate the real data distribution under adversarial conditions. 
The adversary in this case is the discriminator (D) which is a binary classifier whose inputs are either samples from the real distribution, or \emph{fake} samples made up by G. 
The training process is a game between G and D: G is trying to fool D to accept its outputs as \emph{real}, and D gets better in detecting fake inputs from G and distinguishing them from real data. As a result, G adjusts its parameters to move towards the real data manifold described by the training data \cite{goodfellow2014}. 
The described adversarial training can be formulated as the following minmax problem,
\begin{equation}
\min_G\max_D V = ~\mathbb{E}[\log D(\mathbf{x})] + \mathbb{E}[\log (1-D(G(\mathbf{z})))]
\label{eq:GAN_obj}
\end{equation}
\noindent where $V \equiv V(D,G)$ is the value function of the system, referred to as sigmoid cross entropy loss function, 
$\mathbf{x}$ is the feature vector from the real data distribution, 
$\mathbf{z}$ is the latent vector generated from a noisy distribution,
$D(\mathbf{x})$ and $G(\mathbf{x})$ are the outputs of D and G, and $\mathbb{E}$ denotes expected value.

\par In speech enhancement applications, it has been observed that Conditional GAN (CGAN) \cite{mirza2014} results in better performance than conventional GAN \cite{pascual2017,soni2018,pandey2018}. CGAN uses an additional data vector $\mathbf{x_c}$ in both G and D for regression purposes. 
%
Moreover, the GAN method from \eqref{eq:GAN_obj} 
uses sigmoid cross entropy loss function which causes vanishing gradients problem for some fake samples far from the real data, which leads to saturation of the loss function. In the sequel, CGAN is combined with the Least-Squares GAN (LSGAN) \cite{mao2016} which solves this problem by stabilizing GAN training and increasing G's output quality. 
This is achieved by substituting the cross-entropy loss with a binary-coded least-squares function, and training G and D individually. This modified GAN objective function is expressed by,
\begin{equation}
\nonumber
\min_D V(D) =   \mathbb{E}[(D(\mathbf{x},\mathbf{x_c})-1)^2] + \\ \mathbb{E}[(D(G(\mathbf{z},\mathbf{x_c}),\mathbf{x_c}))^2]
\label{eq:lsgan_d}
\end{equation}
\vspace*{-2ex}
\begin{equation}
\min_G V(G) =   \mathbb{E}[(D(G(\mathbf{z},\mathbf{x_c}),\mathbf{x_c})-1)^2]
\label{eq:lsgan}
\end{equation}
 
\vspace*{-1ex}
\section{Proposed System}
\label{sec:sys}
\subsection{Speech Model in the Frequency Domain}
Let $y[m]$ denote the observed noisy speech signal, where $m \in \mathbb{Z}$ is the discrete-time index.
The noisy speech results from the contamination of a desired, 
clean speech signal $s[m]$ with an additive noise signal $n[m]$, i.e.,
\begin{equation}
y[m] = s[m] + n[m], \quad m\in \mathbb{Z}
\label{eq:model_time}
\end{equation}
We represent the signals of interest in the time-frequency domain, as obtained from application of the STFT to \eqref{eq:model_time}. Specifically, the STFT coefficients of the noisy speech signal $y[m]$ are defined as,
\begin{equation}
Y(k,f) = \sum_{m=0}^{M-1} y[m+kL] h[m] e^{-j2 \pi f m /M }
\label{eq:stft}
\end{equation}
where $k \in \mathbb{Z}$ is the frame index,
$L$ is the frame advance,
$f \in \{ 0,1,2,...,M/2\}$ is the frequency bin index,
$M$ is the frame size and
$h[m]$ is a window function.
In practice, the calculation in \eqref{eq:stft} is implemented by means of an $M$-point Fast Fourier Transform (FFT) algorithm.
Applying the STFT formula from \eqref{eq:stft} on the time-domain model \eqref{eq:model_time} yields the  time-frequency model representation
\begin{equation}
Y(k,f) = S(k,f) + N(k,f)
\label{eq:model_stft}
\end{equation}
where $S(k,f)$ and $N(k,f)$ are the STFT of the clean speech and noise signals, respectively.

\subsection{Audio Fingerprinting Features}
\par To train the GAN architecture, we propose a new feature set obtained by combination of MFCC and NSSC. 
In this part, we explain the calculation and combination of these AFP features.
\subsubsection{Mel-Frequency Cepstral Coefficients (MFCC)}
\par MFCC are widely used in speech recognition and enhancement due to their powerful compacting capabilities while preserving essential information in speech \cite{chen2014,razani2017,kolbaek2017}. 
To calculate the MFCC features, the time-domain signal $y[m]$ is passed through a first order FIR filter to boost the highband formants in a so-called pre-emphasis stage, as given by, 
\begin{equation}
    y'[m] = y[m] - \alpha y[m-1]
    \label{eq:preemph}
\end{equation}
where $\alpha$ is the pre-emphasis coefficient, with $0.95 \leq \alpha \leq 1$.
\par Next, the STFT of the filtered signal $y'[m]$ is calculated as in \eqref{eq:stft}, yielding the STFT coefficients $Y'(k,f)$. For each data frame, these STFT coefficients are used to calculate a set of Spectral Subband Energies (SSE) defined in terms of a bank of overlapping narrow-band filters.
Specifically, the SSE of the $k$-th frame are calculated as,
\begin{equation}
    \text{SSE}_y(k,b) = \sum_{f=l_b}^{h_b} w_b(f) |Y'(k,f)|^2 
    \label{eq:sse}
\end{equation}
where $b \in \{0,1,\ldots,B-1\}$, $B$ is the number of subbands in the filterbank,
and $w_b(f) \ge 0$ is the spectral shaping filter of the $b$-th subband, with
$l_b$ and $h_b$ denoting the lower and upper frequency limits of $w_b(f)$.
More specifically, the filters $w_b(f)$ together form a mel-spaced filterbank, i.e., they are characterized by triangular shapes with peak frequencies distributed according to the mel-scale of frequency.
\par Finally, the Discrete Cosine Transform (DCT) - Type III is applied to the logarithm of the SSE to obtain the desired MFCC features, which is expressed as,
\begin{equation}
    \text{MFCC}_y(k,p) = \sqrt{\frac{2}{B}}\sum_{b=0}^{B-1} \log_{10}{\text{SSE}_y(k,b)}\cos{(\frac{p\pi}{B}(b-0.5))}
    \label{eq:mfcc}
\end{equation}
\noindent where $p \in \{0,1,\ldots,P-1\}$ and $P$ is the number of coefficients. 
We define the MFCC feature vector of the current data frame as: $\textbf{MFCC}_y = [\text{MFCC}_y(k,0),...,\text{MFCC}_y(k,P-1)]$.
\subsubsection{Spectral Subband Centroids (SSC)}
\par The SSC were introduced in \cite{paliwal1998} to measure the center of mass of a subband spectrum in terms of frequency, using a weighted average technique. 
These features exhibit robustness against the equalization, data compression and additive noise which do not significantly alter the peak frequencies at moderate to high Signal-to-Noise Ratio (SNR) \cite{duong2015}. 
In \cite{seo2005}, the SSC outperform MFCC when used as inputs in a audio recognition task based on dictionary matching. 
To generate SSC values, the noisy speech signal $y[m]$ is pre-emphasized as in \eqref{eq:preemph} and the corresponding STFT coefficients $Y'(k,f)$ are computed. For each frame, a set of SSC is obtained by calculating the centroid frequencies of a bank of narrowband filters as in the MFCC.
Specifically, the SSC of the $k$-th frame are calculated as,
\begin{equation}
    \text{SSC}_y(k,b) = \frac{\sum_{f=l_b}^{h_b} f \, w'_b(f)|Y'(k,f)|^2}{\sum_{f=l_b}^{h_b}w'_b(f)|Y'(k,f)|^2}
    \label{eq:ssc}
\end{equation}
where $b \in \{0,1,\ldots, B-1\}$ and $w'_b(f)$ is the corresponding subband filter. 
In this work, to simplify implementation, we use the same bank of triangular mel-scale filters for both MFCC and SSC calculations, i.e. $w'_b(f)=w_b(f)$ 
\par Finally, following \cite{seo2005}, the SSC values are normalized within the range $[-1,1]$,
which is more convenient for use in neural network layers and activation functions. The normalized SSC (NSSC) features are obtained as,
\begin{equation}
    \text{NSSC}_y(k,b) = \frac{\text{SSC}_y(k,b) - (h_b - l_b)}{h_b - l_b}
    \label{eq:nssc}
\end{equation}
For later reference, we define the NSSC feature vector of signal $y[m]$ at the current frame $k$ as $\textbf{NSSC}_y = [\text{NSSC}_y(k,0),...,\text{NSSC}_y(k,B-1)]$.
\subsubsection{Feature Combination}
\par 
In this paper, we propose to use the concatenation of MFCC and NSSC vectors, along with some of their first and second differences (i.e., delta and double-delta) for training the GAN architecture. In the sequel, we refer to this extended feature set as AFP Combination (AFPC). 
The MFCC and their deltas have long been used as an efficient alternative to the STFT, as they contain crucial information about the spectral subband energies and their temporal evolution \cite{huang2001}.
Nevertheless, due to the smoothing nature of \eqref{eq:sse}, the MFCC ignore the dynamics of the formant present in each subband.
In contrast, the NSSC and their deltas can provide critical information about the formant locations and their temporal variations. At the same time, the NSSC tend to be more noise-robust, compared to the MFCC, since the formant locations are not significantly disturbed by the additive noise distortion \cite{paliwal1998}. 
Thence, the proposed AFPC features have the ability to capture information about the distribution of energy, both across and inside spectral subbands. 
\par 
To obtain the AFPC, the MFCC and NSSC are both extracted from the STFT of the noisy signal, $Y(k,f)$ as described previously. The proposed AFPC feature vector at the $k$-th time frame for signal $y[m]$ is then defined as, 
\begin{equation}
\begin{split}
\textbf{AFPC}_y =  [\textbf{MFCC}_y, \Delta\textbf{MFCC}_y,\Delta^2\textbf{MFCC}_y,\\ \textbf{NSSC}_y, \Delta\textbf{NSSC}_y, \Delta^2\textbf{NSSC}_y]
\end{split}
\label{eq:afpc_1}
\end{equation}
where $\Delta\textbf{MFCC}_y$ and $\Delta^2\textbf{MFCC}_y$ are the deltas and double-deltas of the MFCC. Similarly, $\Delta\textbf{NSSC}_y$ and $\Delta^2\textbf{NSSC}_y$ are the deltas and double deltas of the NSSC.

\subsection{Incorporation of AFPC within GAN}
\par 
We assume that the magnitude spectrum of the noisy speech can be approximated by the sum of the clean speech and noise magnitude spectra, i.e, $|Y(k,f)| \approx |S(k,f)| + |N(k,f)|$. 
The generator in the adversarial setting is trained to predict a \emph{real} output, which is taken as the Ideal Ratio Mask (IRM) genera-ted from the known clean speech and noise signals  \cite{narayanan2013}, i.e.,
\begin{equation}
\text{IRM}(k,f) = \sqrt{\frac{|S(k,f)|^2}{|S(k,f)|^2+|N(k,f)|^2}} 
\label{eq:irm}
\end{equation}
We define the IRM vector at the current frame $k$ as $\textbf{IRM} = [\text{IRM}(k,0),...,\text{IRM}(k,M/2)]$ . Then, the generator produces the estimated IRM whose patterns and distribution should be close to the real IRM, as expressed by, 
\begin{equation}
    \widehat{\textbf{IRM}} = G(\mathbf{z},\textbf{AFPC}_y^j)
    \label{eq:esirm}
\end{equation}
\noindent where $\textbf{AFPC}^j_y$ represents the AFPC feature vector at the current frame, obtained by concatenating the AFPC feature vectors from a subset of $2j+1$ consecutive context frames centered at the current one (i.e., by including the $j$ adjacent frames to its left and right). The estimated output $\widehat{\textbf{IRM}}$ in \eqref{eq:esirm} is only calculated for the current frame.
\par By examining $\widehat{\textbf{IRM}}$ and the $\textbf{AFPC}_{y}$ of the current frame, D decides whether its input is the real IRM from  \eqref{eq:irm}, or the \emph{fake} output from \eqref{eq:esirm}. The estimated IRM for every frame and frequency index is used as a Wiener type of filter on the STFT magnitude of the noisy speech. This method only enhances the amplitude of the signal and uses the phase from the noisy speech to reconstruct the time-domain enhanced signal using the overlap-add and Inverse STFT (ISTFT) as shown in,
\begin{equation}
|\hat S(k,f)| = \widehat{\text{IRM}}(k,f)|Y(k,f)| 
\label{eq:recon}
\end{equation}
\begin{equation}
\hat s[m] = \text{ISTFT}\{|\hat S(k,f)|e^{jk\angle Y(k,f)}\}
\label{eq:t-recon}
\end{equation}
\par In \cite{pascual2017}, it is reported that having an extra term in training the generator using CGAN is very useful. Pandey \emph{et al.} \cite{pandey2018} show that using the $L_1$ loss gives a better performance compared to the $L_2$ loss in speech enhancement applications. 
This approach allows adversarial component to produce more refined and realistic results. 
The weight of the $L_1$ component in the objective function is controlled by a parameter $\lambda>0$. Therefore, the objective functions from \eqref{eq:lsgan} are modified as,
\begin{equation}
\begin{split}
\min_D V(D) = & ~  \mathbb{E}[(D(\textbf{IRM},\textbf{AFPC}_y)-1)^2] \\ & +\mathbb{E}[(D(G(\mathbf{z},\textbf{AFPC}^j_y),\textbf{AFPC}_y))^2]
\end{split}
\label{eq:se_d}
\end{equation}
\begin{equation}
\begin{split}
\min_G V(G) =  &~ \mathbb{E}[(D(G(\mathbf{z},\textbf{AFPC}^j_y),\textbf{AFPC}_y)-1)^2] \\& + \lambda \| G(\mathbf{z},\textbf{AFPC}^j_y)-\textbf{IRM}\|_1 
\end{split}
\label{eq:se_g}
\end{equation}
\par
\begin{figure}[ht]
\centering
   \includegraphics[scale=0.7] {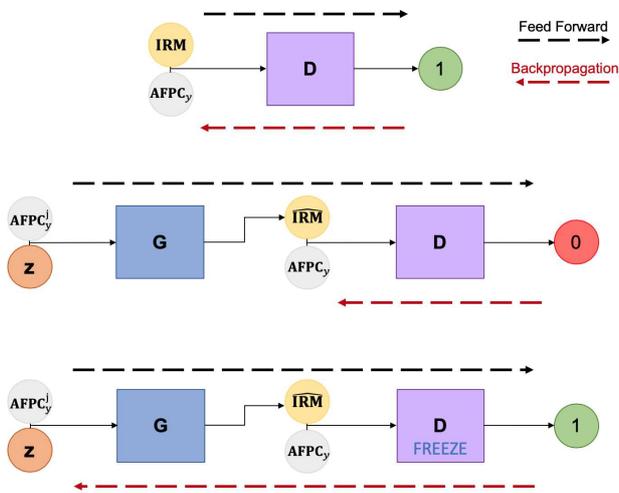}
   \caption{The Proposed GAN training procedure used with the AFPC. 
   } 
   \label{fig:GAN_arch}
\end{figure}

 A schematic of this adversarial training procedure is illustrated in Fig. \ref{fig:GAN_arch}. The training consists of three consecutive steps: 
First, D is trained with a concatenation of the \textbf{IRM} vector and the $\textbf{AFPC}_{y}$ feature vector, in such a way that it recognizes the \textbf{IRM} as real (or output 1). 
Next, D learns to categorize the concatenation of the $\widehat{\textbf{IRM}}$ and $\textbf{AFPC}_y$ feature vector as fake data distribution (or output 0). Finally, D variables are frozen and the G is trained with the $\textbf{AFPC}_y^j$ features to fool the D.  

\par
\begin{figure*}[ht]
\centering
   \includegraphics[scale=0.81] {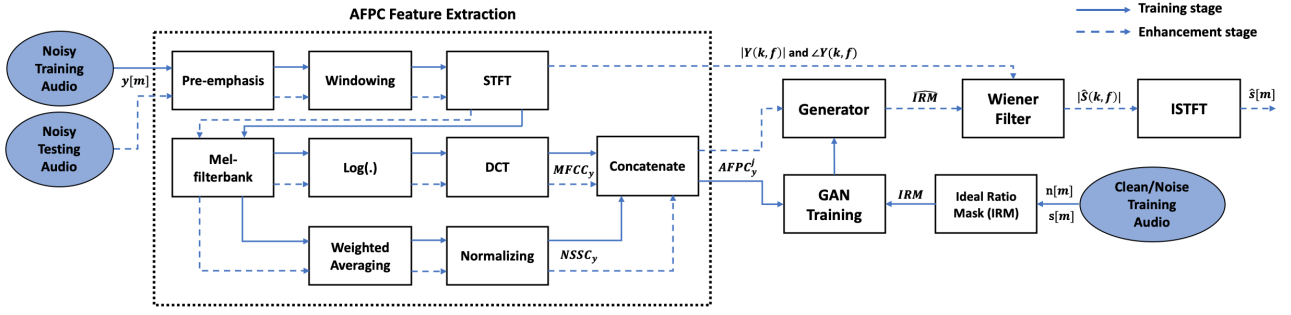}
   \caption{Block diagram of the proposed AFPC training feature set and its incorporation into GAN.} 
   \label{fig:arch}
\end{figure*}

 A block diagram of the system architecture is depicted in Fig. \ref{fig:arch}. The operation consists of two stages: training and enhancement. During the training stage, the system uses the AFPC features to train the D and G as shown in Fig. \ref{fig:GAN_arch} and learn the IRM. In the enhancement stage, the G from the GAN setting is inputted with the AFPC features to output the estimated $\widehat{\text{IRM}}$ and the speech spectrum is reconstructed using a Wiener type of filtering shown in \eqref{eq:t-recon}.

\section{Experimental Setup}
\label{sec:exp}
\subsection{Dataset}
We use the LibriSpeech \cite{panayotov2015} dataset which is an open corpus based on audio books and containing 1000 hours of relatively noise-free speech in English. 
For training, 1755 utterances are randomly selected from 250 speakers (half male, half female) for a total of 6 hours of speech. For testing, 255 utterances are selected from 40 speakers (half male, half female), for a total of 30 minutes of speech. The clean files are contaminated with additive noise at -5dB, 0dB and 5dB SNRs for both training and testing sets, while two extra SNRs of 10dB and 15dB are added for testing. Five different noise types from NOISEX-92 \cite{varga1993} are used for both training and testing: babble, pink, buccaneer2, factory1 and hfchannel. 

\par All the audio files are sampled at 16 KHz. The STFT coefficients are extracted with an $M=512$ STFT, using a 32ms Hanning window, overlap of 50\% ($L=256$) and three context frames (i.e. $j=1$). The MFCC and NSSC are computed from the STFT parameters using $B=64$ subbands with mel-frequency triangular filters $w_b(f)$ distributed between $0$Hz and $8$KHz. 
The number of MFCC is set to $P=22$ while for NSSC, only the first 22 coefficients are kept in the feature vector. The pre-emphasis factor $\alpha = 0.97$ is used in \eqref{eq:preemph}. The delta and double-delta variations are included in the feature sets for each context frame \cite{paliwal1998}. The estimated IRM  \eqref{eq:esirm} is calculated only for the middle STFT frame. For each feature set, one model is trained for all noise types, SNRs and speakers. 

\subsection{Training and Evaluation}
\par The generator's architecture has three hidden layers, each including 512 nodes. The ReLU activation function is used after each hidden layer with a dropout rate of 0.2. The discriminator has the same structure as the generator but uses instead the leaky ReLU activation function. Both employ the sigmoid activation at the output layer because they predict the IRM. The latent vector $\mathbf{z}$ has 15 elements generated randomly from a normal Gaussian distribution. The GAN architecture is trained in 50 epochs with a learning rate of $10^{-4}$ for the first half and $10^{-5}$ for the second half of the epochs. The batch size is set to 128 and ADAM optimizer is used for training. We set $\lambda=100$ in \eqref{eq:se_g}, which provides good convergence.

We compare different combinations of the discussed features, i.e. STFT coefficients, MFCC and NSSC, and they are designated with "+", which means concatenation of the indicated feature vectors. Out of the seven distinct possible combinations, STFT+MFCC+NSSC combination is not included in the study, since it does not substantially improve the performance nor the computational efficiency. In each experiment, one GAN architecture is trained for each feature set using all SNRs and noise types and uses the same architecture, training and hyper-parameters.
The feature sets are compared objectively in terms of PESQ, which provides a measure of signal quality between -0.5 and 4.5, Signal-to-Distortion Ratio (SDR) which measures the speech quality in dB based on the introduced speech distortion, and Short-Time Objective Intelligibility (STOI), which provides a measure of intelligibility between 0 and 1. 
Besides these performance measures, we also compare the different feature combinations in terms of system efficiency, i.e. feature vector size, training time per epoch, and number of network parameters.

\section{Results and Discussion}
\label{sec:res}

\par
\begin{figure}[ht]
\centering
   \includegraphics[scale=0.95] {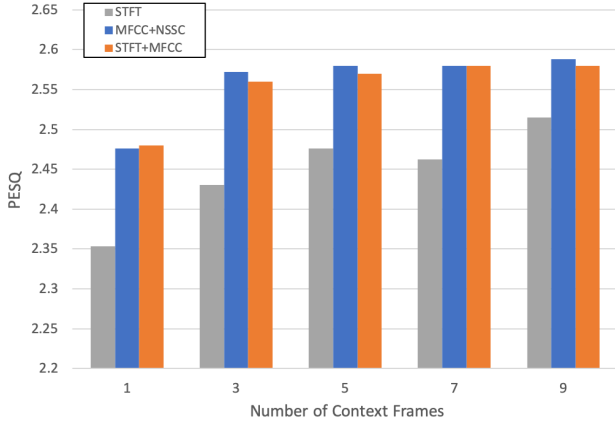}
   \caption{Average PESQ performance for three feature sets: STFT, MFCC+NSSC and STFT+MFCC in different context frames from 1 to 9.} 
   \label{fig:pesq}
\end{figure}

 In this section, we present and discuss the experimental results. To select the number of context frames (i.e., $2j+1$), the PESQ performance of three selected feature sets is studied as demonstrated in Fig. \ref{fig:pesq}. When the number of context frames increases, the performance tend to improve for each  feature set. However, since most of the gains for MFCC+NSSC and STFT+MFCC are obtained with 3 context frames, we use the value of $j=1$ for all subsequent experiments.
\par For each feature set, results are obtained for five different noise types at five SNR levels from -5dB to 15dB. Average PESQ, SDR and STOI measures over all noise types are reported in Tables \ref{tab:ls_pesq}-\ref{tab:ls_stoi}, where the best results (within the 2\% of the observed maximum) are highlighted for each SNR. 
When used separately, MFCC and NSSC improve the overall speech quality compared to the noisy speech but do not generally outperform STFT. 
Comparing STFT with STFT+NSSC and STFT+MFCC indicates that both AFP features add important information to the STFT features. 
STFT+MFCC outperforms STFT+NSSC in terms of both PESQ and STOI, while achieving a similar SDR performance. 
\par According to Tables \ref{tab:ls_pesq}-\ref{tab:ls_stoi}, the proposed AFPC, i.e., MFCC+NSSC, substantially increases the performance of the GAN-based speech enhancement system in all three measures compared to MFCC or STFT. Furthermore, MFCC+NSSC achieves the best PESQ performance (within the error margin) and demonstrates a performance close to STFT+MFCC in terms of SDR and STOI. In particular, MFCC+NSSC outperforms the other feature sets in all three measures at high unmatched SNR of 15dB. This is due to the fact that at such high SNR, the additive noise does not significantly corrupt the extraction of formant frequencies with NSSC.
\begin{table}[ht]
\centering
\caption{Average PESQ Results for all noise types at various SNRs}
\label{tab:ls_pesq}
\begin{tabular}{|c|c|c|c|c|c|}
\hline
\multirow{2}{*}{\textbf{Feature Set}}    & \multicolumn{5}{c|}{\textbf{PESQ}}  \\ \cline{2-6}
				&-5dB&0dB&5dB&10dB&15dB\\ \hline
Noisy 			&1.13&    1.40&    1.72&    2.07&    2.43\\ \hline
STFT  &1.71&2.12&2.52&2.82&2.99 \\ \hline
NSSC  &1.56&2.07&2.48&2.80&3.07 \\ \hline
MFCC  &1.69&2.11&2.50&2.84&3.12 \\ \hline
STFT+NSSC &1.77&2.20&2.60&2.90&3.04 \\ \hline
STFT+MFCC &\textbf{1.83}&\textbf{2.27}&\textbf{2.64}&\textbf{2.94}&3.14 \\ \hline
MFCC+NSSC  &\textbf{1.82}&\textbf{2.25}&\textbf{2.63}&\textbf{2.96}&\textbf{3.21} \\ \hline
\end{tabular}
\end{table}
\begin{table}[ht]
\centering
\caption{Average SDR Results for all noise types at various SNRs}
\label{tab:ls_sdr}
\begin{tabular}{|c|c|c|c|c|c|}
\hline
\multirow{2}{*}{\textbf{Feature Set}}    & \multicolumn{5}{c|}{\textbf{SDR(dB)}}  \\ \cline{2-6}
				&-5dB&0dB&5dB&10dB&15dB\\ \hline
Noisy 	&-5.21&   -0.34&    4.62&    9.61&   14.6 \\ \hline

STFT  &3.80&7.71&11.5&15.1&17.8 \\ \hline
NSSC  &3.05&7.10&10.8&14.0&16.5 \\ \hline

MFCC  &3.17&6.96&10.7&14.3&17.2 \\ \hline
STFT+NSSC &\textbf{4.16}&\textbf{7.95}&\textbf{11.7}&\textbf{15.2}&17.9 \\ \hline
STFT+MFCC &\textbf{4.18}&\textbf{7.96}&\textbf{11.7}&\textbf{15.3}&18.3 \\ \hline
MFCC+NSSC  &\textbf{4.11}&\textbf{7.80}&\textbf{11.6}&\textbf{15.2}&\textbf{18.5} \\ \hline

\end{tabular}
\end{table}
\begin{table}[!ht]
\centering
\caption{Average STOI Results for all noise types at various SNRs}
\label{tab:ls_stoi}
\begin{tabular}{|c|c|c|c|c|c|}
\hline
\multirow{2}{*}{\textbf{Feature Set}}    & \multicolumn{5}{c|}{\textbf{STOI}}  \\ \cline{2-6}
				&-5dB&0dB&5dB&10dB&15dB\\ \hline 
Noisy 	&0.56&0.67&0.78&0.87&0.93 \\ \hline

STFT  &0.69&0.79&\textbf{0.87}&\textbf{0.92}&\textbf{0.94}\\ \hline
NSSC &0.64&0.76&0.85&0.90&0.93\\ \hline
MFCC  &0.68&0.79&0.86&0.91&\textbf{0.94}\\ \hline
STFT+NSSC &\textbf{0.70}&\textbf{0.80}&\textbf{0.88}&\textbf{0.92}&\textbf{0.94} \\ \hline
STFT+MFCC &\textbf{0.71}&\textbf{0.81}&\textbf{0.88}&\textbf{0.92}&\textbf{0.95} \\ \hline
MFCC+NSSC &\textbf{0.70}&\textbf{0.80}&\textbf{0.88}&\textbf{0.92}&\textbf{0.95} \\ \hline
\end{tabular}
\end{table}

\par While the bottom 3 feature sets in Tables \ref{tab:ls_pesq}-\ref{tab:ls_stoi} achieve the best performance in terms of average PESQ, STOI and SDR, the cost of this improvement for a GAN-based system using STFT+NSCC or STFT+MFCC is much more than for the proposed MFCC+NSSC (i.e., AFPC). 
As shown in Table \ref{tab:complex}, the latter significantly outperforms the former in terms of feature size, training time and number of network parameters. 
Specifically, MFCC+NSCC leads to reductions of
59.1\% in memory storage for the training data, 
43.3\% in training time for the GAN system,
and 25.0\% in the number of network parameters. Compared to the STFT baseline, MFCC+NSCC requires 49.6\% less memory storage for features 
and 30.1\% less training time, while achieving 
significant performance improvements. 
The savings in training time and network size with the proposed AFPC become larger when we add more context frames (i.e., $j>1$). 
The testing time is not reported in Table IV since it is almost the same for all systems. In testing, most of the processing time is allocated to the STFT computation which is similar for all feature combinations.

\begin{table}[ht]
\centering
\caption{Feature Vector Size and Training Time per Epoch}
\label{tab:complex}
\begin{tabular}{|c|c|c|c|c|}
\hline
\textbf{\textbf{Feature Set}} &
\makecell{\textbf{Average} \\ \textbf{PESQ}} &
\makecell{\textbf{Feature} \\ \textbf{Size}} & \makecell{\textbf{Training Time} \\ \textbf{per epoch}} & \makecell{\textbf{Network} \\ \textbf{Param.}} \\ \hline
STFT	&	2.43	& 257 & 17.6 mins & 1.06M\\ \hline
STFT+NSSC	&2.50		& 323& 21.7 mins & 1.16M\\ \hline
STFT+MFCC& \textbf{2.56}		& 323& 21.7 mins & 1.16M\\ \hline
\textbf{MFCC+NSSC} & \textbf{2.57} & \textbf{132}& \textbf{12.3 mins} & \textbf{870K}\\ \hline
\end{tabular}
\end{table}

\vspace{5mm}
\par
\begin{figure}[ht]
\scriptsize
 \begin{tabular}{cc}
(a) Clean speech & (b) Noisy speech \\
\includegraphics[width=40mm]{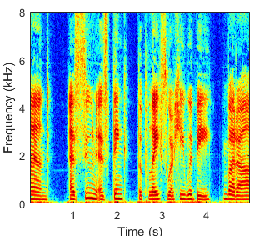}&
\includegraphics[width=40mm]{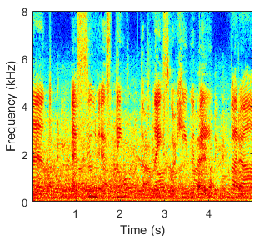}\\
(c) Processed with STFT & (d) Processed with AFPC\\
\includegraphics[width=40mm]{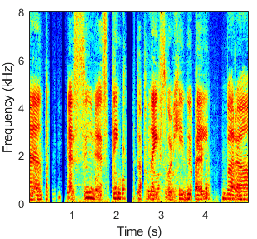}&
\includegraphics[width=40mm]{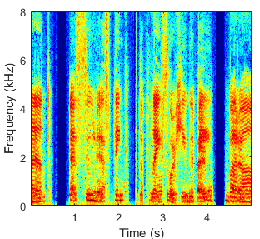}

  \end{tabular}
   \caption{(a) Clean speech (b) Noisy speech (0dB babble noise) (c) Processed speech using STFT features (d) Processed speech using the AFPC features.} \label{fig:spec}

\end{figure}

 Fig. \ref{fig:spec} shows the spectrograms of: (a) clean speech; (b) noisy speech after contamination with babble noise at 0dB SNR; (c) enhanced speech using GAN with STFT, and; (d) enhanced speech using proposed AFPC.
It can be seen that the proposed AFPC features preserve the speech formants while removing more noise during non-speech segments.

\section{Conclusion}
\label{sec:con}
In this work, we proposed using a compact set of features obtained from the combination of two AFP techniques, i.e., MFCC and NSSC, to implement a speech enhancement system based on GAN and trained to predict the IRM of the noisy speech. 
The NSSC capture the speech formants and the distribution of energy in each subband, and therefore complement the MFCC in a crucial way.
In experiments with diverse speakers and noise types,  GAN-based  speech  enhancement  with  the  proposed AFPC (MFCC+NSCC) achieved the best average performance in terms of PESQ, STOI and SDR objective measures.
Furthermore, compared to the STFT+MFCC combination with nearly similar performance, 
AFPC led to reductions of about
60\% in memory storage, 
45\% in training time,
and 25\% in network size.
Hence, the proposed AFPC set is a promising feature-extraction method in learning-based speech enhancement systems.






\begin{thebibliography}{99}
\bibitem{benesty2005}
J. Benesty, S. Makino, J. D. Chen, Speech Enhancement, Berlin, Germany:Springer, 2005.
\bibitem{xu2015} 
Y. Xu, J. Du, L. R. Dai and C. H. Lee, "A regression approach to speech enhancement based on deep neural networks," in \emph{IEEE/ACM Trans. on Audio, Speech, and Language Proc.}, vol. 23, no. 1, pp. 7-19, Jan. 2015.
\bibitem{goodfellow2014}
I. Goodfellow, J. Pouget-Abadie, M. Mirza, B. Xu, D. Warde- Farley, S. Ozair, A. Courville, and Y. Bengio, "Generative adversarial nets, in advances in neural information processing systems," NIPS, Montreal, Canada, 2014, pp. 2672-2680.
\bibitem{pascual2017}
S. Pascual, A. Bonafonte, and J. Serr, "Segan: Speech enhancement generative adversarial network," \emph{Interspeech}, Stockholm, Sweden, 2017, pp. 36423646.
\bibitem{donahue2017}
C. Donahue, B. Li, and R. Prabhavalkar, "Exploring speech enhancement with generative adversarial networks for robust speech recognition," arXiv:1711.05747, 2017.
\bibitem{soni2018}
M. H. Soni, N. Shah and H. A. Patil, "Time-frequency masking-based speech enhancement using generative adversarial network," in \textit{Proc. ICASSP}, Calgary, Alberta, Canada, 2018, pp. 5039-5043.
\bibitem{pandey2018}
A. Pandey and D. Wang, "On adversarial training and loss functions for speech enhancement," in \textit{Proc. ICASSP}, 2018, p. in press.
\bibitem{abdul2019}
J. Abdulbaqi, Y. Gu, I. Marsic, "RHR-Net: A residual hourglass recurrent neural network for speech enhancement," arXiv:1904.07294, 2019.
\bibitem{ribas2019}
D. Ribas, J. Llombart, A. Miguel, and L. Vicente, "Deep speech enhancement for reverberated and noisy signals using wide residual networks," arXiv:1901.00660, 2019.
\bibitem{razani2017}
R. Razani, H. Chung, Y. Attabi, B. Champagne, "A reduced complexity MFCC-based DNN approach for speech enhancement", in \emph{Proc. IEEE Symp. on Signal Process. and Information Tech.}, pp. six, Dec. 2017.
\bibitem{chen2014} 
J. Chen, Y. Wang, and D. Wang, "A feature study for classification-based speech separation at low signal-to-noise ratios," in
IEEE/ACM Trans. on Audio, Speech, and Language Processing, vol. 22, no. 12, pp. 1993–2002, 2014.
\bibitem{duong2015}
N.Q. Duong and H.T. Duong, "A review of audio features and
statistical models exploited for voice pattern design," arXiv preprint arXiv:1502.06811. 2015 Feb 24. 
\bibitem{paliwal1998}
K. K. Paliwal, "Spectral subband centroids features for speech recognition,"
in \textit{Proc. ICASSP}, vol. 2, Seattle, U.S, 1998, pp. 617–620.
\bibitem{poh2004}
N. Poh, C. Sanderson, and S. Bengio, “An investigation of spectral subband centroids for speaker authentication,” in \emph{Proc. LNCS} 3072, \emph{Int. Conf. Biometric Authent. (ICBA)}, Hong Kong, 2004, pp. 631–639
\bibitem{nicolson2018}
A. Nicolson, J. Hanson, J. Lyons, and K. Paliwal, "Spectral subband centroids for robust speaker identification using marginalization-based missing feature theory," \emph{Int. Journal of Signal Processing Systems}, vol. 6, no. 1, pp. 12–16, 2018.
\bibitem{Kinnunen2007}
T. Kinnunen, B. Zhang, J. Zhu, and Y. Wang, "Speaker verification with adaptive spectral subband centroids," Adv. Biometrics, pp. 58–66, 2007.
\bibitem{thian2004}
N. Thian, C. Sanderson, and S. Bengio, "Spectral subband
centroids as complementary features for speaker authentication," Biometric Authenticat., pp. 1-38, 2004
\bibitem{mirza2014}
M. Mirza and S. Osindero, "Conditional generative adversarial nets," arXiv:1411.1784, 2014.
\bibitem{mao2016}
X. Mao, Q. Li, H. Xie, R. Y. K. Lau, and Z. Wang, “Least squares generative adversarial networks,” arXiv: 1611.04076, 2016.
\bibitem{kolbaek2017} M. Kolb\ae k, Z. Tan and J. Jensen, "Speech intelligibility potential of general and specialized deep neural network based speech enhancement systems," in IEEE/ACM Trans. on Audio, Speech, and Language Processing, vol. 25, no. 1, pp. 153-167, 2017.
\bibitem{seo2005}
J. S. Seo, M. Jin, S. Lee, D. Jang, S. Lee, C. D. Yoo, "Audio fingerprinting based on normalized spectral subband centroids," in \emph{Proc. ICASSP}, Philadelphia, USA, vol. 3, pp. 213-216, Mar. 2005.
\bibitem{narayanan2013}
A. Narayanan and D. L. Wang, "Ideal ratio mask estimation using deep neural networks for robust speech recognition," in \emph{Proc. ICASSP}, Vancouver, Canada, 2013, pp. 7092–7096.
\bibitem{huang2001}
X. Huang, A. Acero, and H. Hon, Spoken language processing: a guide to theory, algorithm, and system development, Prentice Hall, 2001.
\bibitem{panayotov2015}
V. Panayotov, G. Chen, D. Povey and S. Khudanpur, "Librispeech: An ASR corpus based on public domain audio books," in \emph{Proc. ICASSP}, Brisbane, Australia, 2015, pp. 5206-5210.
\bibitem{varga1993}
A. Varga, H. J. M. Steeneken, "Assessment for
automatic speech recognition: II. NOISEX-92: A database and an experiment to study the effect of additive noise on speech recognition systems," Speech Communication, vol.12, no.3, pp. 247-252, 1993.

\end{thebibliography}
\end{document}